\RequirePackage{fix-cm}
\pdfoutput=1
\documentclass[global,twocolumn]{svjour}
\smartqed  
\usepackage{graphicx}
\usepackage{gensymb}
\usepackage{graphics}
\usepackage{cite}
\usepackage{textgreek}
\usepackage[T1]{fontenc}

\journalname{myjournal}
\newcommand{\textss}[1]{\scriptsize \mbox{#1}}
\newcommand{\units}[1]{\mbox{ }\mbox{#1}}

\def\funding{\par\addvspace{17pt}\small\rmfamily
	\begin{trivlist}\if!\fundname!\item[]\else
	\item[\hskip\labelsep
	{\it\fundname}]\fi}
\def\endfunding{\end{trivlist}\addvspace{6pt}}
\newcommand\fundname{Funding\runinend}

\begin{document}
\title{Tunable visible frequency combs from a Yb-fiber-laser-pumped optical parametric oscillator}
\author{Yuning Chen\inst{1}\thanks{Y. Chen and M.C. Silfies contributed equally to this work.} \and Myles C. Silfies\inst{1}$^\star$ \and Grzegorz Kowzan\inst{2} \and Jose Miguel Bautista\inst{1} \and Thomas K. Allison\inst{1}\thanks{Corresponding author: thomas.allison@stonybrook.edu}
}                     
\institute{Departments of Chemistry and Physics, 100 Nicolls Rd, Stony Brook University, Stony Brook, NY, 11794, USA
	\and
	Institute of Physics, Faculty of Physics, Astronomy and Informatics, Nicolaus Copernicus University, ul. Grudzi\k{a}dzka 5/7, 87-100 Toru\'n, Poland
}
\date{Received: date / Revised version: date}
%
\maketitle
\begin{abstract}
We present a 100 MHz repetition rate Yb-fiber-laser-pumped synchronously pumped optical parametric oscillator (SPOPO) delivering tunable frequency combs covering almost the entire visible spectral range. By intracavity doubling both the signal and idler combs and using collinear residual pump light, nearly continuous tuning over the range of 420-700 nm is achieved with only small gaps near the OPO degeneracy condition. Output powers range from 10 mW to 200 mW, depending on wavelength, with pulse durations below 150 fs without external compression. Frequency locking of all three collinearly outcoupled combs (pump, doubled signal, and doubled idler) to a femtosecond enhancement cavity facilitates direct comparison of their optical phase noise and phase modulation transfer functions. In the singly-resonant OPO, optical phase modulation of the pump comb is transferred nearly completely to the non-resonant idler comb. This results in a resonant signal comb with reduced optical phase noise and also enables high-bandwidth modulation on the idler comb via phase modulation of the pump. 

\end{abstract}
\section{Introduction}
Frequency comb lasers have achieved tremendous success in many applications, at many different wavelengths \cite{Newbury_NatPhot2011,Cingoz_Nature2012,Timmers_Optica2018,Quinlan_RSI2010,Yang_Optica2016}. In metrology applications where a frequency comb is used to measure the frequency of a continuous wave laser or calibrate a spectrograph, supercontinuum generation in highly nonlinear fibers (HNLF) usually provides sufficient power at the wavelength to be measured. For applications in which the frequency comb is used to interact with atoms and molecules directly, i.e. ``Direct-frequency comb spectroscopy" (DFCS) \cite{Adler_AnnRevChem2010,Coddington_Optica2016,Weichman_JMolSpec2019}, higher spectral brightness than is available from fiber supercontinua is desired. Thus, the generation of tunable or broadband high-power frequency combs has recently been a subject of intense study, particularly in the mid- and long-wave infrared \cite{Timmers_Optica2018,Schliesser_NatPhot2012,Seidel_SciAdv2018,LeinDecker_OptExp2012,Lee_OptLett2013,Ruehl_OptLett2012,Sobon_OptLett2017,Steinle_OptLett2016,Steinle_OptExp2014}.

In this article, we address the problem of tunable visible frequency comb generation, which has received somewhat less attention. Our target application is widely tunable cavity-enhanced ultrafast transient absorption spectrosopy (CE-TAS), recently demonstrated at one wavelength by our research group \cite{Reber_Optica2016}. A tunable CE-TAS system operating in the visible can record the ultrafast dynamics of electronically excited jet-cooled molecules \cite{Barbatti_PCCP2009}, clusters \cite{Kungwan_PCCP2012}, and radicals \cite{Vansco_JChemPhys2017} in the same way that conventional ultrafast spectrometers routinely exploit this spectral region to study the dynamics of condensed phase samples \cite{Berera_PhotoSynthRes2009}. This is a demanding application, requiring high powers ($>$ 10 mW), short pulses, control over the comb's carrier-envelope offset frequency ($f_0$), low optical phase noise, and a wide tuning range due to the broad spectral features inherent to molecules undergoing ultrafast dynamics. A robust fiber-based backbone, requiring minimal maintenance, is also desired due to the complexity of the downstream experiment. 

Previous work has reported the achievement of subsets of this list. Doubling the tunable dispersive waves from an Er:fiber laser pumped HNLF in a periodically-poled lithium niobate crystal (PPLN), Moutzouris et al. \cite{Moutzouris_OptLett2006} demonstrated multi-millliwatt tunable comb generation at 108 MHz repetition rate. However, the pulse durations were long (300-1000 fs) and the tuning range only extended down to 520 nm. In another example, using a series of HNLFs pumped with 1020 nm laser pulses, Tu et al. \cite{Tu_OptExp2009} demonstrated dispersive wave generation in the spectral range of 347 - 680 nm. This covered nearly the entire visible with average powers of 1-9 mW, but to cover the whole tuning range, 7 different HNLFs were required, necessitating replacing the fiber and realigning the system to change wavelength. Using fiber-laser-pumped synchronously pumped optical parametric oscillators (SPOPO), widely tunable combs with larger average powers \cite{Tian_OptLett2016,Cleff_ApplPhysB2011} and shorter pulses \cite{Coluccelli_OptLett2017} have been achieved. Some work has explored intracavity doubling and sum frequency generation for tunable visible and UV comb generation \cite{Gu_OptLett2013,Gu_OptExp2015,Fan_OptExp2016,McCracken_OptLett2015}. However, these previous fiber-laser-pumped SPOPOs have only  demonstrated coverage of small portions of the visible spectrum. SPOPOs pumped by the second harmonic of Ti:Sapphire lasers have achieved impressive tuning ranges \cite{Ghotbi_OptLett2006} covering nearly the entire visible with the signal beam alone. Such combs can in principle be well stabilized \cite{Sun_OptLett2007}, but still require managing the drawbacks of the Ti:Sapphire pump comb compared to fiber lasers, namely greater expense, higher maintenance, and faster $f_0$ drifts when not actively stabilized.

In this article, we report the development of a Yb-fiber-laser-pumped singly-resonant SPOPO which delivers short pulses, high-power, and low optical phase noise throughout the visible spectral range. The OPO cavity is designed to intracavity double both the signal and idler beams using a single BBO crystal, and also utilize the collinear and co-polarized pump comb. In this way, nearly continuous tuning over the range of 420-700 nm ($>$ 9000 cm$^{-1}$) is achieved with only small gaps near OPO degeneracy, where signal and idler wavelengths are equal. Frequency locking of all three collinearly outcoupled combs (pump, doubled signal, and doubled idler) to a $\sim$100 kHz linewidth femtosecond enhancement cavity \cite{Jones_PRA2004} facilitates direct comparison of their optical phase noise and phase modulation transfer functions. We find that the phase relations established for singly resonant OPOs \cite{Kobayashi_JOpt2015,Kobayashi_OptLett2000} cause phase modulation to be transferred nearly completely from the pump to the non-resonant idler. More precisely, the phase modulation of the pump acts on the resonant signal comb with a fixed point \cite{Newbury_JOSAB2007,Telle_AppliedPhysB2002} near the signal comb optical carrier frequency. This has important consequences for the noise and stabilization of the combs. 

While our target application has been widely tunable CE-TAS, this work can also be useful to researchers working on more conventional ultrafast spectroscopy, high-resolution multidimensional spectroscopy \cite{Lomsadze_Science2017}, visible dual-comb spectroscopy \cite{Bergevin_NatCom2018}, ultrafast nonlinear microscopy \cite{Liebel_NatPhot2017,Polli_LaserPhotRev2018}, and researchers working on the stabilization of OPOs in general \cite{Kobayashi_JOpt2015}. 

\section{Experimental setup}\label{sec:setup}
\begin{figure*}
 \includegraphics[width = \textwidth]{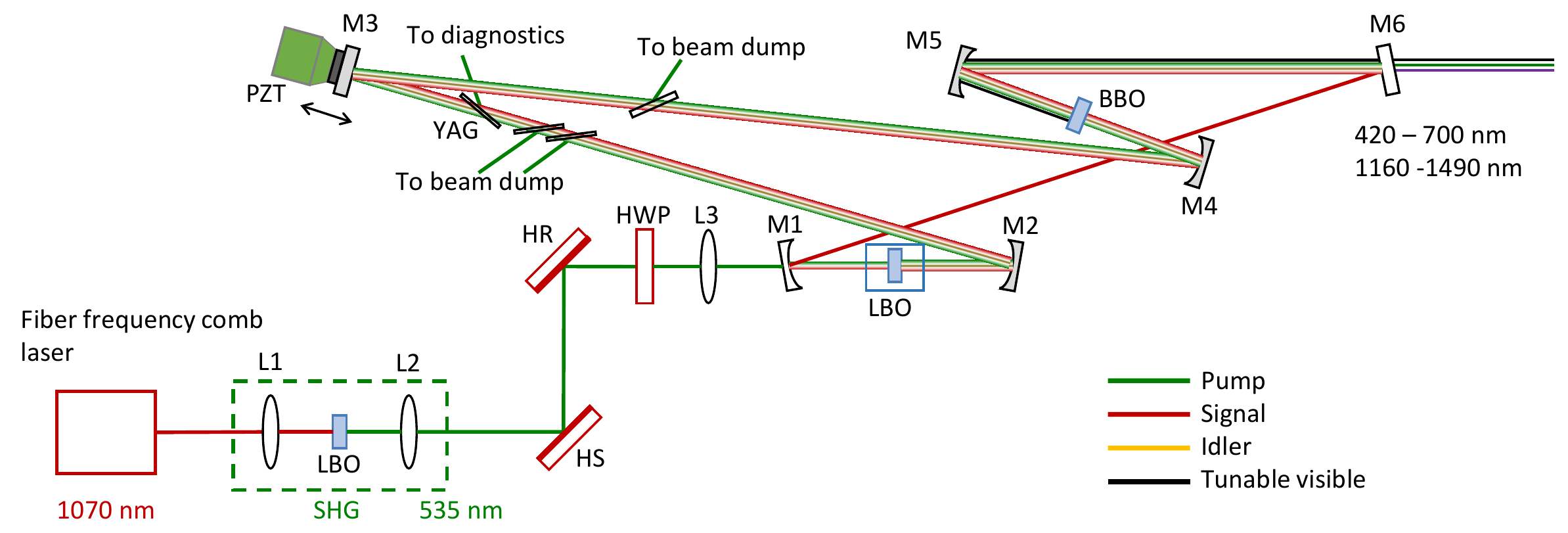}
 \caption[The optical layout of the synchronously pumped optical parametric oscillator.] {Schematic of the synchronously pumped optical parametric oscillator. A frequency comb laser centered at 1070 nm is frequency doubled in an LBO crystal to 535 nm, providing the pump for the OPO. The tunable visible light is generated by  frequency doubling both the signal and idler. Mirror M5 collimates the visible light and M6 couples the light out of the cavity. L1, L2: focal length = 10 cm; HS: harmonic separator; HR: high reflector; HWP: half waveplate; L3: focal length = 15 cm; M: mirror. More details in the main text.}
\label{fig:OPOLayout}
\end{figure*}

The optical layout of the OPO is shown in Fig.~\ref{fig:OPOLayout}. The pump frequency comb at 535 nm is initially derived from a commercial mode-locked Er:fiber laser (Menlo Systems Ultra Low Noise variant), supplying a frequency comb at 1560 nm. The Er:fiber comb is shifted to 1.07 $\mu$m using a combination of a nonlinear Er-doped fiber amplifier and propagation in a short HNLF in similar manner to Maser et al. \cite{Maser_ApplPhysB2017}. The shifted comb is then amplified in a large-mode area photonic crystal fiber amplifier previously described in \cite{Li_RSI2016}. After the compressor, the pulses are 120 fs long with a pulse energy of 115 nJ at a repetition rate of 100 MHz. This laser is frequency doubled in a 2 mm thick LBO crystal cut for type I second harmonic generation (SHG) at 1064 nm with a conversion efficiency of 40\%. The  resulting 4.5 watts at 535 nm is the pump of the OPO.

The OPO cavity contains two nonlinear crystals, a lithium triborate (LBO) crystal for parametric gain and a beta barium borate (BBO) crystal for intracavity doubling the signal and idler. The LBO crystal is 3 mm long and cut for type I noncritical phase-matching. The crystal has angles of $\theta$=90\degree \space and $\phi$=0\degree \cite{Chen:1989tm}. Both sides have anti-reflection coatings at 535 nm and 850 - 1060 nm for the pump and signal respectively. The crystal is housed in a compact home-built oven for controlling the crystal temperature between 25\degree C and 200\degree C \cite{Cleff_ApplPhysB2011}. The compact oven design allows small ($\sim$ 90 mrad angle of incidence) angles on the curved mirrors M1 and M2 for reducing astigmatism and the OPO's spatial footprint.

Intracavity second harmonic generation of both signal and idler is phase matched by rotating the BBO crystal about it's horizontal axis ($\theta$). We select BBO for the broad phase-matching tunability and moderately high nonlinear coefficient. Using SNLO \cite{SNLO:um}, we calculate the phase-matching angle for type I SHG of 850 nm is $\theta =$ 27.6\degree \space and SHG of 1400 nm is $\theta = $ 20.0\degree. The BBO crystal has a cut angle of $\theta$ = 19.8\degree \space for type I SHG of 1500 nm and is protective coated. BBO crystals of both 1 and 2 mm length were used, with some results for both shown in Sect. \ref{sec:results}. The doubled output is optimized by tuning $\theta$ while adjusting the OPO cavity length to maintain resonance.

The OPO cavity is designed to be resonant for the signal from 850 nm to 1060 nm, focus both the signal and idler at the BBO crystal, and collimate and outcouple the doubled idler ($2i$), doubled signal ($2s$), and residual pump in a collinear and co-polarized fashion. This is achieved by using protected silver mirrors for M2-M5 and dichroic mirrors ($R > 99.9$\% at 850 - 1060 nm and  $R < 5$\% reflectivity for 450 - 700 nm). All four curved mirrors have a radius of curvature of 20 cm. The two dichroic mirrors M1 and M6 are lossy for the idler, so that resonance is only attained for the signal. The round trip loss of the signal power caused by the optics is estimated to be 16\%, resulting in a finesse of $\approx 35$. The pump is vertically (s) polarized, and thus the signal and idler are generated at the LBO crystal with horizontal (p) polarization. The doubled signal and doubled idler are then generated with vertical s-polarization in the BBO crystal, such that the pump, $2s$, and $2i$ beams emerge collimated, collinear, and co-polarized from M6.

Four YAG plates at Brewster's angle are placed in the cavity to attenuate the residual s-polarized pump light at the BBO crystal and the output coupler (M6) to ~200 mW. YAG was chosen for it's high index of refraction and corresponding shallow Brewster's angle. The YAG plates also provide intracavity dispersion enabling stable operation closer to degeneracy. One of the YAG plates is mounted on a rotation stage so it can be rotated from the Brewster condition by a known angle, allowing us to outcouple a fraction of the signal and idler to diagnostics, the fraction calculated from the Fresnel equations. The outcoupled light is analyzed using calibrated Germanium photodetectors and an optical spectrum analyzer.

We control the OPO cavity length by moving mirror M3 with a micrometer and two piezoelectric transducers (PZT). The micrometer is used for coarsely finding the resonance condition of OPO free spectral range $=\textrm{ pump } f_{\textss{rep}}$. A long travel ($\sim$ 10 micron) PZT stage allows further coarse tuning of the OPO cavity length and long-term in-loop correction for drift. A faster ``copper bullet" style PZT \cite{Briles_OptExp2010} is used for frequency stabilization of the OPO when using the doubled signal beam. The incident angle on M3 is set to be $\sim$ 1.5\degree \space, so that the cavity misalignment is negligible when changing the cavity length on the sub-mm scale. 

\begin{figure*}
\begin{center}
 \includegraphics[width = 0.9\textwidth]{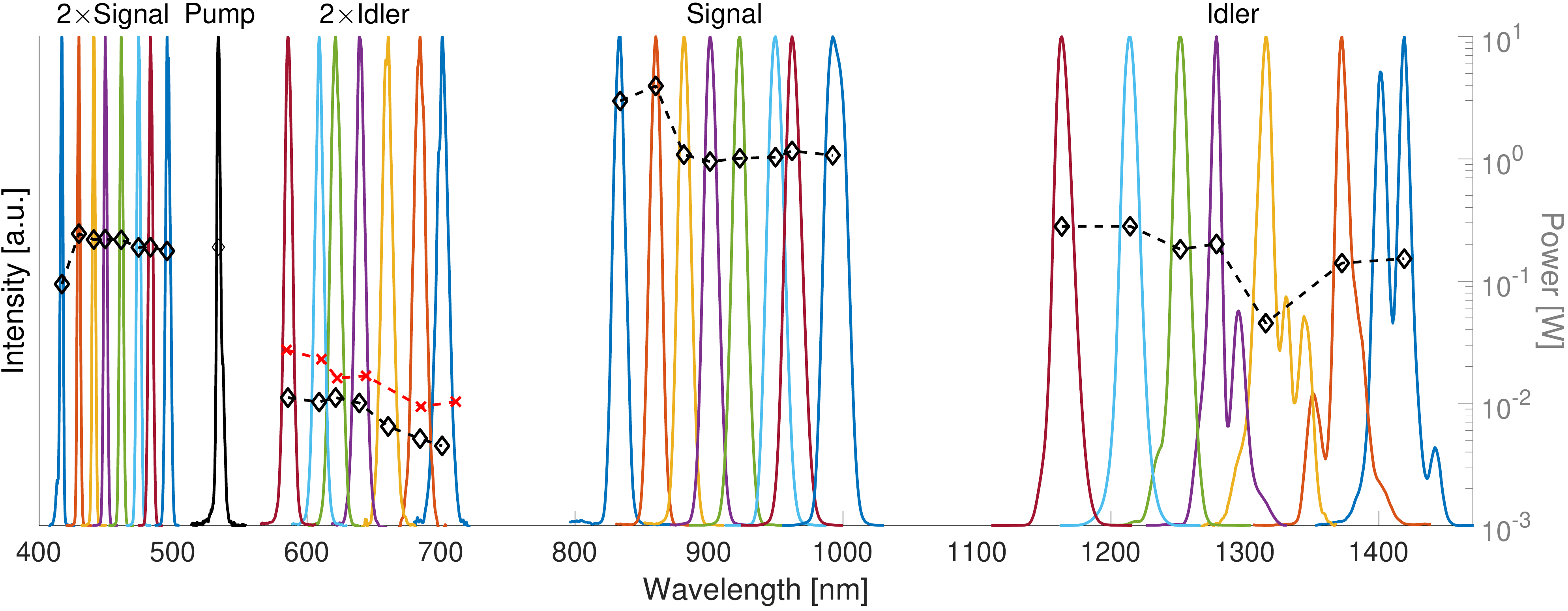}
 \caption{Intracavity signal and idler spectra and power at different phase-matching temperatures of LBO in the synchronously pumped optical parametric oscillator. The spectrum centered at 535 nm is the residual pump. Spectra for fundamental and second harmonic are color matched. Optical powers are represented by diamonds for 1 mm BBO and crosses for 2 mm BBO ($2i$ only) and use the right axis. Reported powers for pump, $2i$, and $2s$ are OPO output while signal and idler are intracavity power levels.}
  \label{fig:SPOPOSpectraPower}
\end{center}
\end{figure*}

\section{Results}\label{sec:results}

The threshold pump power required to see parametric oscillation was measured to be around 1.7 W, and was not strongly affected by the insertion of the doubling crystal. The spectrum and power of signal and idler with 4.5 watts pump power are shown on the right side of Fig.~\ref{fig:SPOPOSpectraPower}. The signal and idler intracavity powers are reported with the diamond symbols using the right axis of the figure. By varying the LBO temperature from 122\degree C to 145\degree C and adjusting the OPO cavity length as in \cite{Cleff_ApplPhysB2011}, the signal is tuned between 835 nm to 990 nm and the corresponding idler ranges from 1490 nm to 1160 nm. Tuning towards shorter signal wavelengths is limited by the coating of M1 and M6. As we get closer to degeneracy, where signal and idler wavelengths are equal, the signal and idler spectra become broader and unstable due to the low intracavity dispersion and the broad phase matching bandwidth in LBO. The signal and idler spectra displayed in Fig.~\ref{fig:SPOPOSpectraPower} were taken when their respective second harmonics were optimized. Removing the BBO typically resulted in roughly a factor of 2 increase in intracavity power for both signal and idler. 

\begin{figure}[t]
\includegraphics[width=\linewidth]{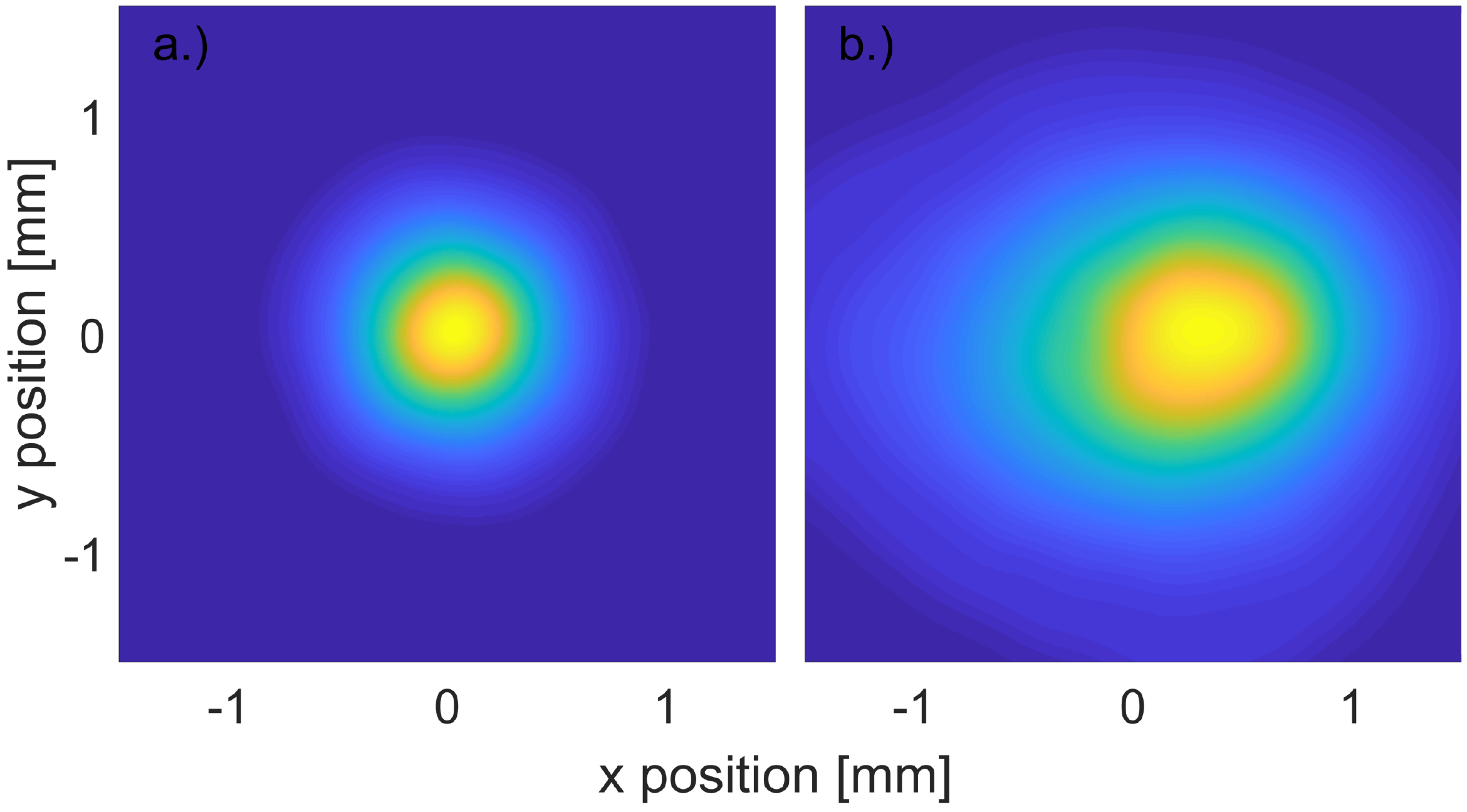}
\caption{Output spatial modes for, a.) the doubled signal and, b.) the doubled idler. Gaussian fits to the vertical and horizontal profiles for each give beam waists of $\omega_x = 313$ \textmu m and $\omega_y = 316$ \textmu m for the doubled signal and $\omega_x = 547$ \textmu m and $\omega_y = 479$ \textmu m for doubled idler.}
\label{fig:modes}
\end{figure}

The spectra and output power of the tunable visible combs provided by the combination of $2s$, $2i$, and pump are also shown in Fig.~\ref{fig:SPOPOSpectraPower}, along with outcoupled power using the right axis. The displayed spectra were taken with the 1 mm BBO in the cavity. With the 1 mm thick BBO crystal the outcoupled $2s$ power is 100-200 mW and the outcoupled $2i$ power is 5-10 mW across the tuning range. Figure \ref{fig:modes} shows the spatial profiles of the outcoupled $2s$ and $2i$ beams using the 1 mm BBO crystal for doubling. With the 2 mm thick BBO crystal, higher doubled powers are attained but the spatial profile of the $2s$ beam degrades significantly, while beam quality of the $2i$ beam was retained. Power with the longer crystal is also plotted for the doubled idler in Fig.~\ref{fig:SPOPOSpectraPower}, indicated by cross marks and using the right axis. 

\begin{figure}
    \centering
    \includegraphics[width = \linewidth]{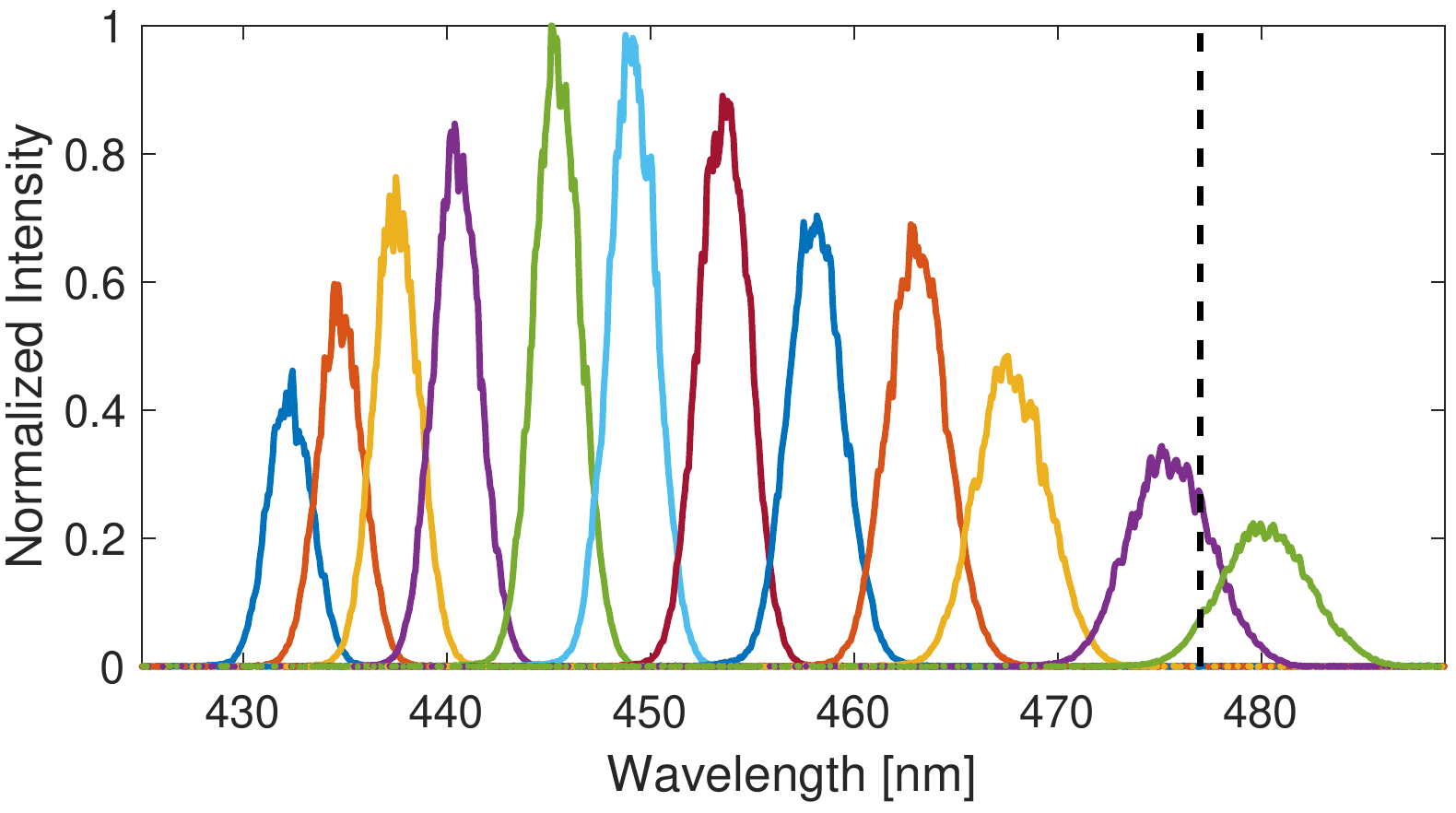}
    \caption{Doubled signal output spectra at a single LBO crystal temperature, T = 138\degree \space C as the OPO cavity length and BBO angle are tuned. All spectra normalized to the highest signal observed in the spectrometer. Dashed vertical line indicates the signal wavelength corresponding to the optimum doubled idler center wavelength for the same temperature.}
    \label{fig:SignalAtOneTemp}
\end{figure}

When doubling the idler, the idler wavelength found by optimizing the $2i$ power agree with both calculations for the LBO phase matching for the given crystal temperature as well as the idler wavelength found by optimizing the intracavity idler power without the BBO crystal. However, the oscillating signal wavelength found by optimizing $2s$ power is shifted, as illustrated in Fig.~\ref{fig:SignalAtOneTemp}. The LBO parametric gain is phase matched over a large bandwidth due to the low group-velocity walkoff in the LBO crystal \cite{SNLO:um}, and the OPO can thus oscillate far away from optimum parametric gain. Doubling the resonant signal influences the OPO oscillation and causes the optimum $2s$ power to be found blueshifted from the wavelength expected from both the LBO phase matching calcuation and the optimum idler/$2i$ wavelengths found when not doubling the signal. Additional parasitic or cascaded nonlinear processes are not observed, and the idler wavelength shifts accordingly when the doubled signal was optimized. 

\begin{figure}[t]
 \includegraphics[width = \linewidth]{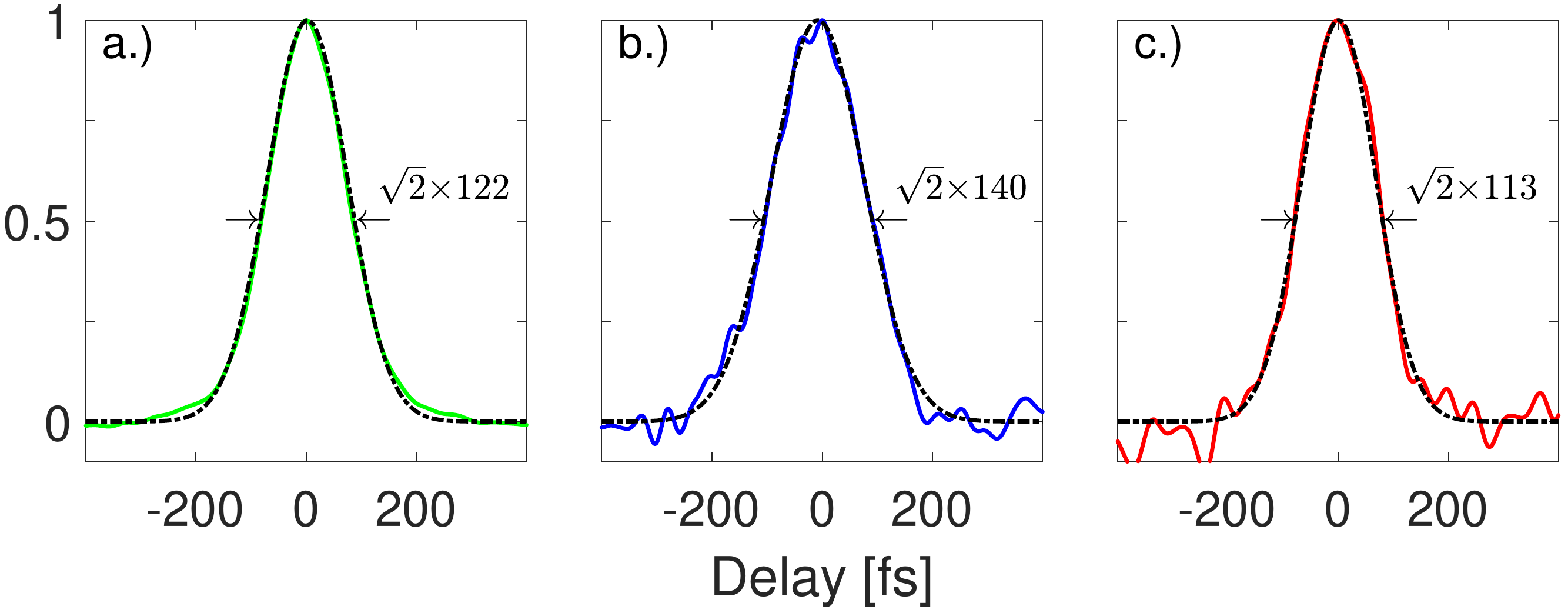}
 \caption{Normalized intensity autocorrelation traces for different visible outputs from SPOPO (solid) and Gaussian fit (black, dashed): a.) residual pump, b.) doubled signal, c.) doubled idler. Quoted pulse durations are FWHM assuming Gaussian pulse shape.}
 \label{fig:OPOPulseDuration}
\end{figure}

We measured the visible pulse durations with a home-built two-photon autocorrelator. The interferometric autocorrelator uses an all-reflective design similar to the one reported in \cite{Mashiko_ApplPhysB2003} and uses a SiC two-photon photodiode (IFW Optronics) for the detector. The data is low-pass filtered to show only the intensity autocorrelation signal. Autocorrelations for the pump, $2s$, and $2i$ wavelengths are shown in Fig.~\ref{fig:OPOPulseDuration} as well as Gaussian fits to the data. Assuming Gaussian pulses, all pulse durations are less than 150 fs.

\begin{figure}
\includegraphics[width = \linewidth]{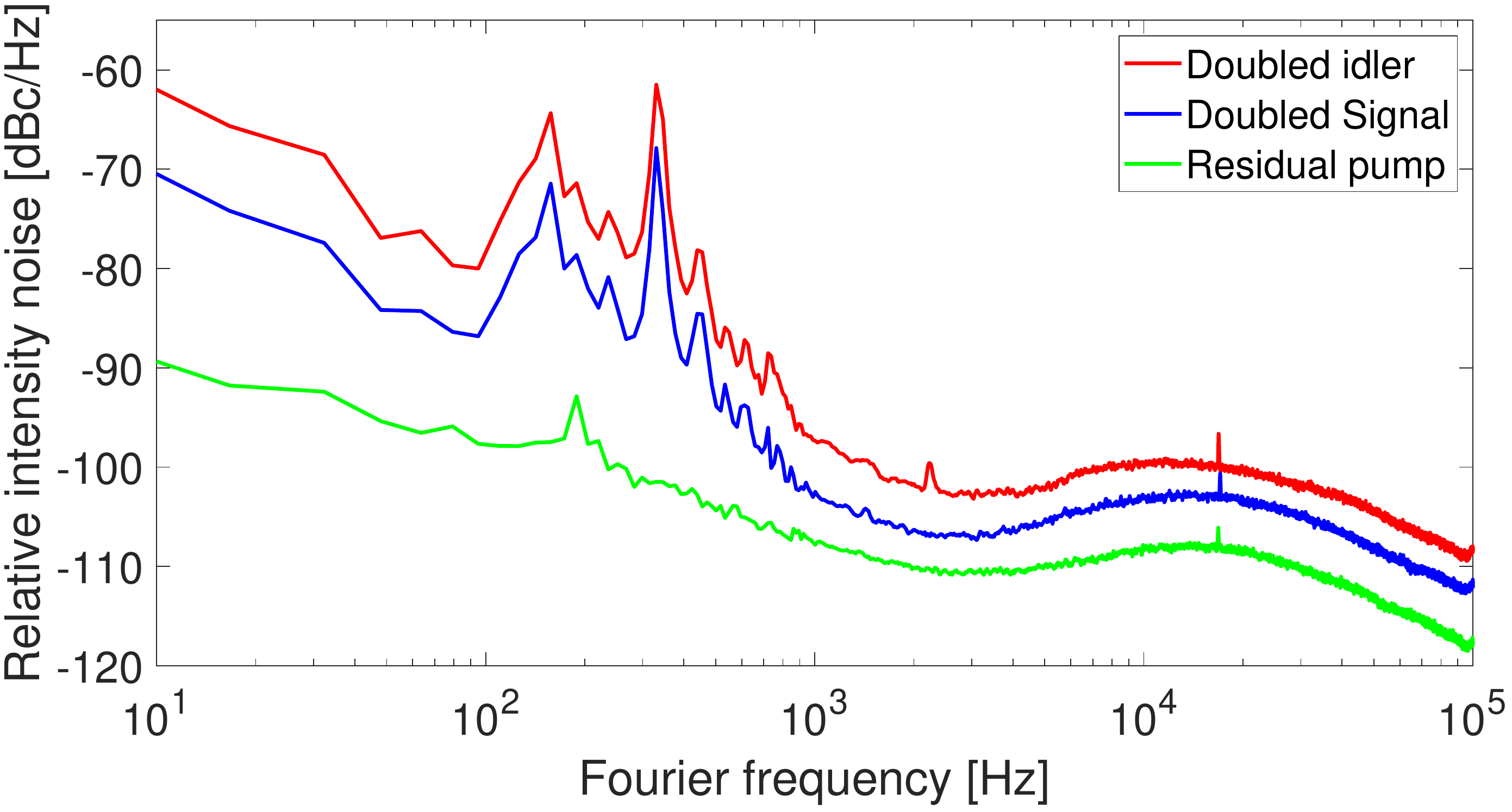}
 \caption{Relative intensity noise (RIN) of OPO pump, doubled signal, and doubled idler.}
 \label{fig:FreeRIN}
\end{figure}

\begin{figure}
\includegraphics[width = \linewidth]{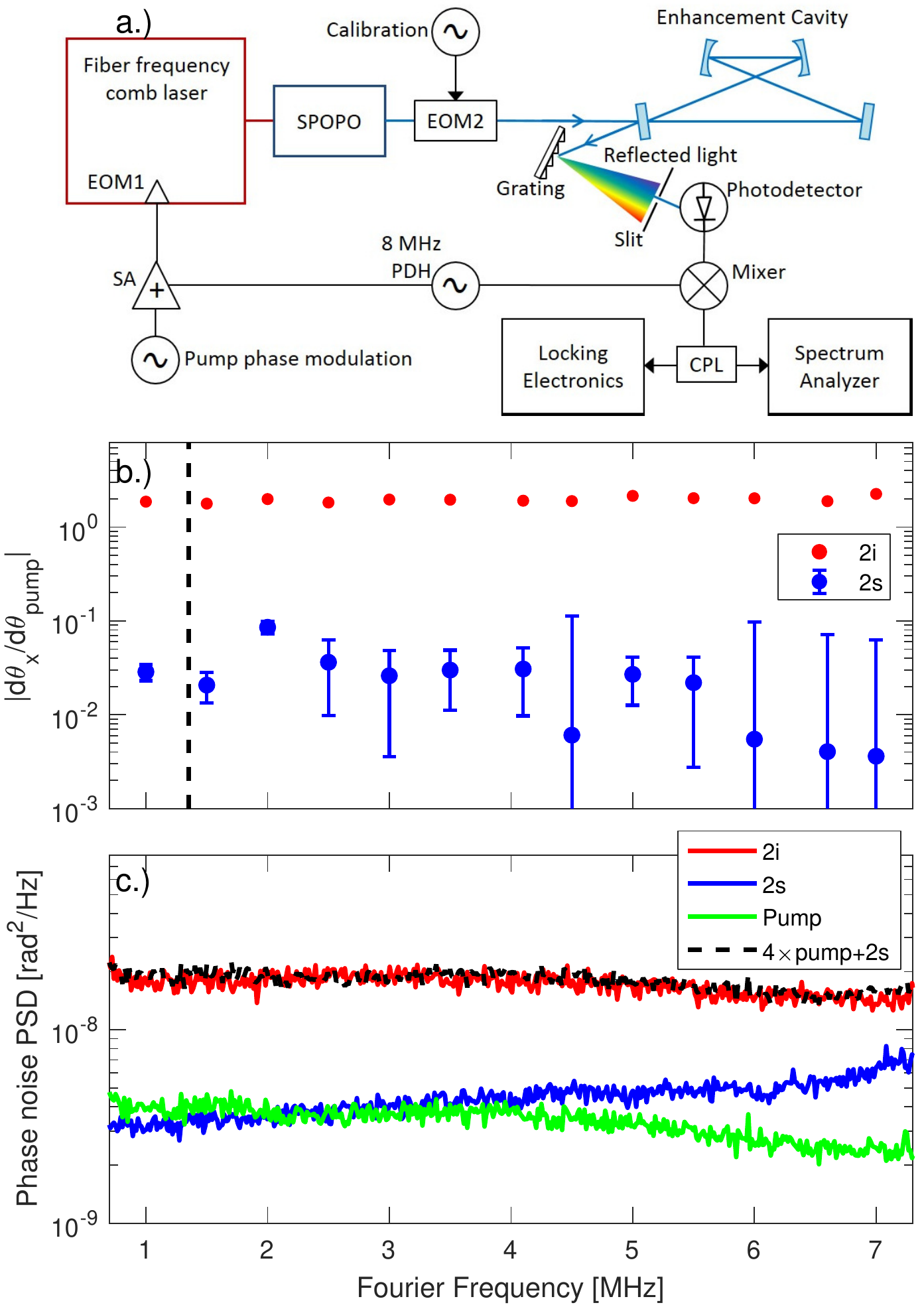}
 \caption{a.) Setup for measuring pump $\rightarrow 2i$ transfer function or phase noise of pump or $2i$ combs. See text for details as well as changes required for measuring $2s$ comb. SA: summing amplifier, CPL: directional coupler. b.) Pump $\rightarrow 2i/2s$ phase modulation transfer functions. Dashed vertical line indicates OPO cavity half-linewidth. c.) High frequency phase noise power spectral density of pump, doubled signal and doubled idler measured as well as the sum of  Eq.~(\ref{eqn:phase_combine2}) for comparison to $2i$ noise.}
 \label{fig:phasenoise}
\end{figure}

For our target application of cavity-enhanced ultrafast spectroscopy \cite{Reber_Optica2016}, both low optical phase noise and low amplitude noise (RIN) are critical. In Figs. \ref{fig:FreeRIN} and \ref{fig:phasenoise} we report both. For the RIN plot, the signal is recorded with a home-built low-noise amplified Si detector. At high frequencies, the RIN of the $2s$ and $2i$ beams is comparable to the low-RIN fiber laser used as the pump and competitive with high-performance fiber laser combs \cite{Wunram_OptLett2015}. At frequencies below 1 kHz, the greater susceptibility of the free-space OPO cavity to acoustic and mechanical perturbations is seen in both the $2s$ and $2i$ RIN spectra.

For phase noise, the situation is more interesting. In steady state, the three frequency combs must share the same repetition rate and the optical pulses must obey the phase relationship \cite{Shen_book1984,Kobayashi_OptLett2000,Kobayashi_JOpt2015}
\begin{equation}\label{eqn:phaserel}
	\theta_p = \theta_s + \theta_i + \pi/2
\end{equation}
where $\theta_s$, $\theta_i$, and $\theta_p$ correspond to the optical carrier phases (i.e. carrier-envelope offset phase) of the signal, idler, and pump pulses, respectively. This relationship was explicitly verified by Kobayashi and Torizuka \cite{Kobayashi_OptLett2000} for a singly-resonant OPO by observing hetrodyne beats between the doubled signal and pump + idler. The beat notes were found to shift as the OPO cavity length was changed, determining the relation
\begin{equation}\label{eqn:dids}
	\left( \frac{\partial \theta_i}{\partial \theta_s} \right)_p = -1
\end{equation}
The notation $()_p$ is used to indicate that no actuation is done on the pump comb. The pump comb's $f_0$ is fixed and the relation shows direct phase transfer from signal to idler. 

If instead the optical phase of the pump comb is perturbed, either via $f_{\textss{rep}}$ or $f_0$, in principle the signal and idler can share this phase change via Eq.~(\ref{eqn:phaserel}) in a non-trivial way. To study how pump phase is shared between signal and idler, we measure the pump $\rightarrow 2i,2s$ phase modulation transfer function and phase noise using the setup shown in Fig.~\ref{fig:phasenoise}a. The combs are stabilized to a 4-mirror femtosecond enhancement cavity (fsEC) with a finesse of $\sim 1000$ and net intracavity GDD $<$ 100 fs$^2$ using a two-point Pound-Drevel-Hall (PDH) technique \cite{Drever_ApplPhysB1983,Jones_PRA2004}. For locking the pump and doubled idler, 8 MHz PDH sidebands and fast feedback are applied to the combs via an electro-optic modulator, EOM1, in the Er:fiber oscillator, as shown in Fig.~\ref{fig:phasenoise}a. For the doubled signal, since the phase modulation transfer from the pump to the doubled signal is near zero (shown below), PDH sidebands are applied using a second EOM2 placed between the SPOPO and the fsEC, and fast feedback is applied to the copper bullet PZT on M3 of the OPO cavity. More technical details of the two-point locking feedback loops, broadband comb/cavity coupling, and the attainable performance are being prepared for publication in a forthcoming paper \cite{Silfies_InPrep}. In the present discussion, we use the cavity as an optical phase discriminator. At high frequencies well above both the fsEC linewidth and the frequency-lock servo bandwidth, the PDH error signal acts as a phase discriminator \cite{Zhu_JOSAB1993,Nagourney_Book2010} and is insensitive to the details of the fsEC linewidth or the servo-loop. Essentially, the light reflected from the cavity is heterodyned with the intracavity light. A grating is used to disperse the spectrum reflected from the fsEC \cite{Jones_PRA2004}, so only a small fraction of the frequency comb bandwidth contributes to the PDH error signal used to record phase modulation.

To measure the pump $\rightarrow 2i,2s$ phase modulation transfer functions, we apply sinusoidal phase modulation to the pump via EOM1 and record the modulation strength transferred to the $2s$ and $2i$ combs via the modulation sideband strength observed in the PDH error signal, similarly to \cite{bienfang_phase_2001}. The doubled signal and idler data is then divided by the the pump phase modulation strength measured the same way. The resulting pump $\rightarrow 2i,2s$ transfer function is shown in Fig.~\ref{fig:phasenoise}b. The mean of the doubled idler transfer data (red) is 1.96 and the standard deviation is 0.13. The signal transfer coefficient (blue) is more than 20 times smaller than pump $\rightarrow 2i$ at all modulation frequencies. Since $\theta_{2i} = 2\theta_{i}$, the data in Fig.~\ref{fig:phasenoise}b imply

\begin{equation}\label{eqn:didp}
	\left( \frac{\partial \theta_i}{ \partial \theta_p} \right)_{\ell} = \frac{1}{2}\left( \frac{\partial \theta_{2i}}{\partial \theta_p} \right)_{\ell} = 1
\end{equation}

\begin{equation}\label{eqn:dsdp}
	\left( \frac{\partial \theta_s}{\partial \theta_p} \right)_{\ell} \approx 0
\end{equation}
The notation $()_{\ell}$ is used to indicate that no actuation is done on the SPOPO cavity length. To our knowledge this is the first explicit verification of relations (\ref{eqn:didp}) and (\ref{eqn:dsdp}) and direct measurement of the transfer functions in Fig.~\ref{fig:phasenoise}b for a singly resonant OPO. This un-equal sharing result is quite different than the degenerate case measured by Wan et al. \cite{Wan_OptLett2018}, where phase modulation on the pump is shown to be shared equally between the degenerate subharmonic signal and idler combs. The transfer function in Fig.~\ref{fig:phasenoise}b also shows that the OPO cavity linewidth is irrelevant for the transfer of optical phase from the pump to the non-resonant comb since there is no observable roll-off at the SPOPO cavity half-linewidth of 1.35 MHz.  

Since all three combs must share the same $f_{\textss{rep}}$ (at least for modulation frequencies below the SPOPO cavity linewidth), Eq.~(\ref{eqn:dsdp}) implies that for fixed SPOPO cavity length small repetition rate changes of the pump comb $\Delta f_{\textss{rep}}$ are counterbalanced by $f_0$ changes of the signal comb order $\Delta f_{0,s} = -\Delta f_{\textss{rep}} \frac{\nu_s}{f_{\textss{rep}}}$, where $\nu_s$ is the signal's optical carrier frequency. In this way, the fixed point \cite{Newbury_JOSAB2007} of the signal comb under pump actuation is the signal's optical carrier frequency. More precisely, the data of Fig.~\ref{fig:phasenoise}b constrain the location of this fixed point to be within 15 THz of the signal's optical carrier frequency.

We also used the setup of Fig.~\ref{fig:phasenoise}a to measure the optical phase noise of the $2s$, $2i$, and pump combs. Figure \ref{fig:phasenoise}c shows the measured optical phase noise power spectral densities for Fourier frequencies $1 \units{MHz}  < f < 7 \units{MHz}$ in rad$^2$/Hz. The y-axis of the figure is calibrated by applying a known phase modulation to all three combs in identical fashion using EOM2. The pump and doubled signal are at similar levels while the doubled idler is about 4 times (6 dB) higher. This can be understood using the phase transfer relations (\ref{eqn:dids}), (\ref{eqn:didp}), (\ref{eqn:dsdp}). Equations (\ref{eqn:didp}) and (\ref{eqn:dsdp}) imply that optical phase fluctuations of the pump are transferred with unit efficiency to the idler, and any intrinsic OPO noise on the signal comb is uncorrelated. This implies that the two noise sources contribute to the idler phase noise in quadrature, viz.
\begin{equation}\label{eqn:phase_combine1}
    (\Delta \theta_{i})^2 = \left(\frac{\partial \theta_i}{\partial \theta_p} \right)_{\ell}^2 (\Delta \theta_{p})^2 + \left(\frac{\partial \theta_i}{\partial \theta_s}\right)_p^2 (\Delta \theta_{s})^2
\end{equation}
or in terms of the $2i$, $2i$ and pump combs
\begin{equation}\label{eqn:phase_combine2}
    (\Delta \theta_{2i})^2 = 4\left(\frac{\partial \theta_i}{d \theta_p} \right)_{\ell}^2 (\Delta \theta_{p})^2 + \left(\frac{\partial \theta_i}{\partial \theta_s}\right)_p^2 (\Delta \theta_{2s})^2
\end{equation}
The sum on the right-hand side of Eq.~(\ref{eqn:phase_combine2}) is also plotted on Fig.~\ref{fig:phasenoise}c as a dashed, black line, for comparison with the the measured $2i$ phase noise power spectral density. Excellent agreement is found over the entire measured frequency range. 

\section{Conclusions}

In this article, we have provided a detailed description of the design and performance of a fiber-laser-pumped OPO delivering high-power frequency combs which are tunable across the visible region. The high output power and broad tuning range make the combs particularly suitable for performing ultrafast nonlinear spectroscopy with combs, where frequency comb methods have recently allowed large improvements in resolution \cite{Lomsadze_Science2017,Ideguchi_Nature2013} and sensitivity \cite{Reber_Optica2016}. The tunable combs can also be used in other comb applications such as high-resolution cavity-enhanced (linear) comb spectroscopy \cite{Hoghooghi_Optica2019,Adler_AnnRevChem2010}, or in myriad less specialized applications where tunable high-repetition rate femtosecond sources are needed \cite{Liebel_NatPhot2017,Polli_LaserPhotRev2018}.

 Using an external cavity as a phase discriminator, we directly compared the phase noise of all three combs and the pump $\rightarrow$ signal and pump $\rightarrow$ idler phase modulation transfer functions. We find that phase perturbations on the pump are transferred to the (non-resonant) idler comb via the optical phase relations (\ref{eqn:didp}) and (\ref{eqn:dsdp}). These relations have two important consequences for the use of the SPOPO's signal and idler beams for comb applications. First is that the phase noise of the resonant signal is uncorrelated to that of the pump. This further helps explain the extremely low noise performance obtainable with (resonant) OPO combs, in addition to the technical reasons discussed in \cite{Kobayashi_JOpt2015}. The second is that actuation on the non-resonant SPOPO-generated comb via the phase of the pump does not suffer a low-pass filter due to the SPOPO cavity. This enables the use of high-bandwidth transducers, such as EOMs, in the pump comb's mode-locked laser cavity \cite{Benko_OptLett2012,Li_RSI2016} to be used to control the phase of the non-resonant comb (in our case the idler) of the OPO without needing to consider the SPOPO cavity linewidth in the design. For our SPOPO, with $\sim 2.7$ MHz linewidth, the SPOPO cavity would not impose a significant low-pass filter in any event, but this point is important for SPOPOs that use higher cavity finesses to achieve low pump-power thresholds for oscillation \cite{Ebrahimzadeh_ApplPhysB2001}.

\begin{funding}
	National Science Foundation (NSF) (1708743)
\end{funding}
 
\begin{acknowledgement}
M.C.~Silfies acknowledges support from the GAANN program of the U.S. Dept. of Education. G.~Kowzan acknowledges support from the National Science Centre, Poland scholarship 2017/24/T/ST2/00242. We thank S.A.~Diddams, H.~Timmers, A.~Kowligy, A.~Lind, F.C.~Cruz, N.~Nader, and G.~Ycas for helpful discussions and assistance with Er comb development.
\end{acknowledgement}

\end{document}